\documentclass[letterpaper]{article} % DO NOT CHANGE THIS
\usepackage[preprint]{aaai2027}  % preprint option removes copyright for arXiv
\usepackage[hyphens]{url}  % DO NOT CHANGE THIS
\usepackage{graphicx} % DO NOT CHANGE THIS
\usepackage{natbib}  % DO NOT CHANGE THIS AND DO NOT ADD ANY OPTIONS TO IT
\usepackage{caption} % DO NOT CHANGE THIS AND DO NOT ADD ANY OPTIONS TO IT
\usepackage{algorithm}
\usepackage{algorithmic}

\usepackage{newfloat}
\usepackage{listings}
\usepackage{amsmath,amssymb}
\usepackage{multirow}
\DeclareCaptionStyle{ruled}{labelfont=normalfont,labelsep=colon,strut=off} % DO NOT CHANGE THIS
\floatstyle{ruled}
\newfloat{listing}{tb}{lst}{}
\floatname{listing}{Listing}

\usepackage{booktabs}

\title{Independent Patch Verification for Coding Agents with a Bidirectional Reconstruct-and-Verify Framework}
\author{
    Chenglin Li\textsuperscript{\rm 1}\equalcontrib,
    Yisen Xu\textsuperscript{\rm 1}\equalcontrib,Zehao Wang\textsuperscript{\rm 1},Shin Hwei Tan\textsuperscript{\rm 1},Tse-Hsun (Peter) Chen\textsuperscript{\rm 1}
}
\affiliations{
    \textsuperscript{\rm 1}Concordia University, Montreal, QC, Canada
    chenglin.li@mail.concordia.ca,
yisen.xu@mail.concordia.ca, w\_zeha@encs.concordia.ca,
shinhwei.tan@concordia.ca, peterc@encs.concordia.ca
}

\begin{document}

\maketitle

\begin{abstract}
Autonomous coding agents powered by large language models can now generate code patches directly from bug reports, but a fundamental gap remains: once a patch is produced, no mechanism independently verifies whether it truly resolves the reported problem. Prior work has sought to address this through iterative self-refinement and inference-time scaling, but these approaches either review the patch under the same interpretation that produced it or broaden candidate generation without verifying individual patches, and neither provides an explicit verification signal for assessing patch correctness. We propose RETRACE, a training-free post-generation verification framework that derives such a signal through bidirectional reconstruction and reconciliation. When a coding agent generates a candidate patch for an issue, RETRACE performs forward reconstruction to build an explicit repair rationale from the issue and the agent's trajectory; backward reconstruction then independently infers, from the patch and its trajectory alone and without access to the original issue, a description of the problem the patch appears to address, and compares this reconstruction against the original issue to produce an alignment verdict; a reconciliation stage then checks the consistency between the forward rationale and the patch, diagnoses the source of any misalignment, and either submits the patch or produces targeted revision guidance. Evaluated on SWE-bench Verified with two backbones (GPT-5-mini and MiniMax-2.5), RETRACE lifts Pass@1 by 7.0 and 3.6 percentage points respectively on the mini-SWE-agent scaffold, and delivers comparable gains on OpenHands without modification. Ablation experiments show that both the forward and backward stages contribute to the overall improvement and that adding reconciliation yields further gains. Comparison with Self-Refine confirms that the improvement stems from the verification signal and targeted revision guidance rather than from additional computation alone.
\end{abstract}

%     % \link{Datasets}{https://aaai.org/example/datasets}
%     % \link{Extended version}{https://aaai.org/example/extended-version}

% can not use input to include the tex file.
\section{Introduction}
Large language model (LLM) agents have made rapid progress on autonomous software issue resolution, in which the task is to produce a correct code patch from a natural-language bug report and a repository snapshot. Recent work has advanced this capability along multiple fronts: richer agent-environment interfaces and unified code-action paradigms \cite{yang2024sweagent,openhands}, fixed localize-repair-validate pipelines \cite{xia2024agentless}, structure-aware and knowledge-graph-guided code search \cite{zhang2024autocoderover,ma2025kgexplore}, ensemble-based reasoning that decomposes resolution into generation, pruning, and selection \cite{gao2025ensemble}, and inference-time scaling through repeated sampling, trajectory replay, repair plan generation, and evolutionary refinement  \cite{ehrlich2025scaling,niels2025swereplay,li2026historicalpatchesrepairplans,zeng2025evolutionary}. 

Despite this progress, a fundamental limitation of the single-pass pipeline persists: after producing a candidate patch, the agent has no independent signal to assess whether the patch actually addresses the reported problem.  A generated patch may remain plausible but incorrect even when it passes the agent's internal validation mechanisms, such as reproduction tests
\cite{li2026benchmarktestsstrongenough}.
Prior work has sought to improve generated patches through additional post-generation deliberation. Iterative self-refinement methods \cite{madaan2023selfrefine,shinn2023reflexion} provide additional reasoning that can improve patch quality, but because the review is conditioned on the same interpretation that produced the patch, it cannot detect errors that stem from that interpretation itself; moreover, even when self-review identifies a potential issue, it lacks a structured diagnosis of \emph{where} the patch diverges from the reported problem, offering generic feedback rather than targeted revision guidance. Inference-time scaling strategies \cite{ehrlich2025scaling,zeng2025evolutionary} take a complementary approach, broadening the search over candidate patches but not still lacking of a signal to assess whether any individual patch correctly addresses the reported problem. To reliably validate a patch, an agent needs an \emph{independent verification signal} that can diagnose how a generated patch diverges from the reported problem and translate that diagnosis into targeted revision guidance.

In this paper, we propose RETRACE, a framework that derives an independent verification signal through bidirectional reasoning and converts detected inconsistencies into targeted revision guidance via repair reconciliation. Its core idea is analogous to verification in mathematics, where a result is checked by reasoning backward from the derived answer to recover the original conditions and assessing whether the two are consistent.
After a candidate patch is produced, RETRACE proceeds in three stages. First, it performs \emph{forward reconstruction} to build an explicit repair rationale from the original issue and the agent’s repair trajectory, capturing the relevant evidence, reasoning process, and intended repair behavior. Second, it independently performs \emph{backward reconstruction} to infer, from the candidate patch and its trajectory, a natural-language description of the issue that the patch appears to address. The reconstructed issue is then compared with the original issue along multiple semantic dimensions to produce an alignment verdict, providing an independent signal for assessing patch correctness. Finally, when the comparison reveals a mismatch, \emph{repair reconciliation} jointly analyzes the forward rationale and backward reconstruction to determine whether the inconsistency originates from the reasoning, the implementation, or both, and generates a targeted revision plan.

We evaluate RETRACE on SWE-bench Verified with two backbones of different capability levels (GPT-5-mini~\cite{openai2025gpt5} and MiniMax-2.5~\cite{minimax2025m25}) using mini-SWE-agent as the primary scaffold. RETRACE raises Pass@1 from 56.2\% to 63.2\% with GPT-5-mini and from 75.8\% to 79.4\% with MiniMax-2.5. On a second scaffold, OpenHands, RETRACE yields gains of comparable magnitude without any modification, raising Pass@1 from 37.5\% to 56.7\% and from 62.5\% to 70.0\% on the two
backbones respectively, confirming that the mechanism is scaffold-agnostic. Ablations show that the forward and backward stages address complementary failure modes and combine for the strongest result. Compared to Self-Refine, which adds comparable inference without independent reconstruction, RETRACE achieves consistently higher Pass@1 across both backbones and scaffolds. Our contributions are as follows:

\begin{itemize}
\item We propose RETRACE, a post-generation verification framework that reconstructs the problem addressed by a candidate patch independently of the original issue and reconciles this reconstruction with an explicit repair rationale to produce targeted revision guidance, providing an independent verification signal without test execution or ground-truth patches.

\item We evaluate RETRACE on SWE-bench Verified across two LLM backbones and two agent scaffolds. The result shows that consistent improvements over  baselines under identical settings, with ablations confirming that the forward and backward stages resolve
complementary sets of issues and combine for the strongest result.

\item Our comparison with Self-Refine shows that comparable additional inference without independent reconstruction does not match RETRACE's gains, confirming that the improvement stems from the verification signal rather than from additional computation alone.
\end{itemize}
% ============================================================================
\section{Related Work}
% ============================================================================

\subsection{LLM-Based Software Issue Resolution}
Large language models have been increasingly applied to resolve real-world software issues. SWE-agent \cite{yang2024sweagent} introduced an agent-computer interface that provides the model with search, navigation, and editing commands for repository interaction.  Agentless \cite{xia2024agentless} showed that a fixed localize-repair-validate pipeline can be competitive without an interactive agent loop. AutoCodeRover \cite{zhang2024autocoderover} adds program-structure-aware search by combining code search APIs with spectrum-based fault localization. OpenHands \cite{openhands} adopts a CodeAct paradigm in which the agent writes and executes Python and bash code as its unified action space. Beyond issue resolution, LLM-based coding agents have been applied to broader software engineering tasks, including repository-level task completion \cite{ni2026gittaskbench}. Recent systems push further with ensemble-based reasoning that decomposes resolution into generation, pruning, and selection \cite{gao2025trae}, comprehensive repository exploration via knowledge graphs \cite{ma2025lingma}, and autonomous repair agents that dynamically plan tool invocations \cite{bouzenia2025repairagent}. These lines largely improve \emph{localization}, \emph{search}, or \emph{sample-and-filter by tests}. 

RETRACE is orthogonal and complementary: it plans only \emph{after} reading the code and, rather than filtering candidates by test execution, verifies a single patch by reconstructing the issue it solves, a signal available even when no reliable test oracle is.

\subsection{Inference-Time Repair Improvement}
Recent work has explored improving resolve rates at inference time without re-training.  One strategy is test-time scaling through repeated sampling: generating multiple candidate trajectories and selecting among them via majority vote \cite{ehrlich2025codemonkeys}, evolutionary refinement \cite{zeng2025satoriswe}, trajectory replay \cite{niels2025swereplay}, distillation-conditioned rollouts \cite{pdr2025}, or . These methods improve resolve rates by broadening the search over candidate solutions, but do not assess whether the selected patch correctly addresses the reported problem.
A related direction trains process reward models to provide finer-grained feedback on agent trajectories \cite{sweprm2025,sweshepherd2025}.  These approaches can detect trajectory-level errors and offer step-level supervision, but require training data derived from execution signals and score trajectories on general quality rather than alignment with the reported problem.
Another strategy is iterative self-refinement of a single solution.  Self-Refine \cite{madaan2023selfrefine} introduces a general framework in which the same model critiques and revises its own output. Reflexion \cite{shinn2023reflexion} extends this by maintaining an episodic memory of verbal feedback from prior attempts. Self-debugging \cite{chen2023selfdebugging} prompts the model to explain its generated code and correct errors based on execution results.  These methods review the output against the original task description, but the review is conducted by the same model that produced the output, without an independent verification signal.

RETRACE differs from all three categories.  First, it targets the quality of a single agent run rather than searching over multiple candidates as in test-time scaling.  Second, it requires no additional training data, unlike learned reward models.  Third, it reconstructs the problem specification from the patch while withholding the original issue, producing an independent verification signal that self-review methods lack, and enabling targeted revision based on where the patch diverges from the intended repair.
\section{RETRACE: Bidirectional Validation}
\label{sec:method}

\begin{figure*}[htbp]
        \centering      \includegraphics[width=\textwidth]{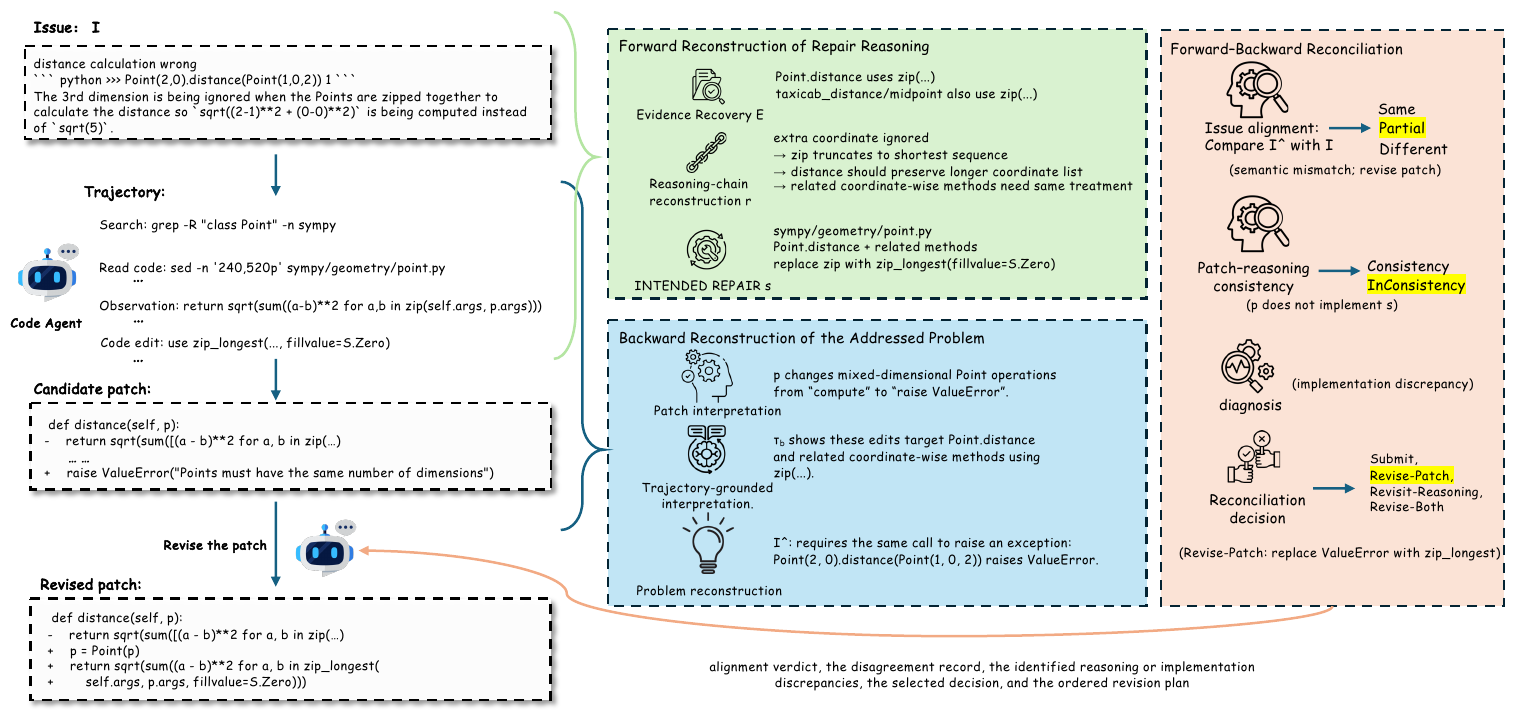}
\caption{Overview of the proposed framework RETRACE}
\label{fig:overview}
\end{figure*}

Figure~\ref{fig:overview} shows the overall workflow of RETRACE. RETRACE is executed after a coding agent produces a candidate patch and combines three stages to generate a verification signal and guide patch revision: (1)~\emph{Forward Reconstruction of Repair Reasoning} derives an explicit repair rationale from the original issue and the agent's trajectory. (2)~\emph{Backward Reconstruction of the Addressed Problem} independently infers the problem addressed by the patch from the implementation and its trajectory. and (3)~\emph{Forward–Backward Reconciliation} compares the two reconstructions against the original issue, and when either check reveals a discrepancy, diagnoses the source of misalignment and produces a targeted revision plan.

Formally, let $I$ denote the original issue, $\tau$ the recorded agent trajectory, and $p$ the candidate patch produced through that trajectory. The patch is treated as the agent's proposed repair rather than as a correct solution. We define two views of $\tau$. The forward view $\tau_{\mathrm{f}}$ contains the repository exploration, observations, and intermediate reasoning available from the trajectory, but not the candidate patch. The backward view $\tau_{\mathrm{b}}$ contains the patch-generation context and repository observations needed to interpret the implementation, but excludes the original issue and explicit restatements of it.

\subsection{Forward Reconstruction of Repair Reasoning}
\label{sec:forward}

For a repair to be reliable, its rationale should be grounded in repository evidence and supported by a logically connected chain of reasoning from the reported issue to the intended code change. However, a trajectory does not necessarily present the agent's decisions as such a chain: relevant evidence may be separated by exploratory detours, intermediate conclusions may be left unstated, and decisions that appear reasonable individually may not support the final repair when considered together. The forward path reconstructs this reasoning into an explicit repair rationale so that RETRACE can examine whether the intended repair follows from the issue and the evidence collected during exploration.

Given the original issue $I$ and the forward trajectory view $\tau_{\mathrm{f}}$, RETRACE derives
\[
\phi = F(I,\tau_{\mathrm{f}})=\bigl(E,r,s\bigr),
\]
where $E$ is the repository evidence recovered from the trajectory, $r$ is the reconstructed reasoning chain, and $s$ is the intended repair. The goal is not to infer a new repair, but to recover the reasoning distributed across the trajectory and determine whether it forms a logically connected path from the reported problem to the intended change.

\paragraph{Evidence recovery.}
The agent may inspect many files, definitions, callers, execution conditions, and tests before reaching a repair. Some observations contribute directly to the final decision, whereas others reflect exploratory detours or abandoned hypotheses. RETRACE traces the decisions leading to $s$ and recovers the repository evidence on which those decisions depend. The resulting evidence set $E$ retains the retains the code and program relationships needed to reconstruct the reasoning while excluding observations that do not contribute to it. Evidence is retained because of its role in the reasoning, not merely because it appears near an edit or shares terminology with the issue. For example, a caller may establish the conditions under which the reported behavior occurs, while a state update may explain why the behavior persists across operations. Conversely, code inspected during an abandoned hypothesis need not be included even when it contains terms from the issue report.

\paragraph{Reasoning-chain reconstruction.}
RETRACE reconstructs an explicit reasoning chain $r$ that links the reported issue to the intended repair using the recovered repository evidence. Rather than replaying the agent trajectory, the reconstruction extracts the key reasoning steps, capturing how the issue is interpreted, which repository evidence supports the diagnosis, and how that evidence justifies the repair direction.

The reconstructed chain is required to satisfy logical consistency. Each reasoning step must be supported by either the reported issue, the selected repository evidence, or a preceding reasoning step, thereby forming a coherent progression from diagnosis to repair. This process distinguishes evidence that substantively contributes to the repair from incidental observations and exposes reasoning gaps where plausible intermediate decisions lack sufficient evidential support or logical justification.

\paragraph{Intended repair.}
The endpoint $s$ describes the repair reached through the reconstructed reasoning, including the behavior to be changed, the relevant implementation location, and the intended scope of the modification. RETRACE does not assume that $s$ is correct. Instead, it determines whether the evidence in $E$ and the reasoning chain $r$ provide a logically connected basis for reaching it.

As the candidate patch $p$ is not available to the forward path, the completed implementation cannot be used to fill missing links or justify earlier decisions after the fact. The resulting reconstruction $\phi$ is subsequently used to determine whether the candidate patch implements the repair established by the agent's reasoning.

\subsection{Backward Reconstruction of the Addressed Problem}
\label{sec:backward}

Beyond sound reasoning, a reliable repair requires that the resulting patch
remain aligned with the originally reported problem. In practice, the candidate patch may not fully preserve the problem interpretation or repair intent developed earlier in the trajectory. It may implement only part of the intended change, extend beyond its intended scope, or modify behavior associated with a related problem. Evaluating the patch directly against the original issue can obscure these deviations because the issue itself supplies the interpretation under which the patch is reviewed. RETRACE instead first reconstructs the problem expressed by the implemented change and only then compares that reconstruction with the reported problem.

Let $\tau_{\mathrm{b}}$ denote the backward view of the agent trajectory, containing the repository observations and implementation context used to produce the patch, but excluding the original issue $I$ and explicit restatements of it. Given the candidate patch $p$ and $\tau_{\mathrm{b}}$, RETRACE derives
\[
\widehat{I}=B(p,\tau_{\mathrm{b}}),
\]
where $\widehat{I}$ describes the problem implied by the implemented changes and their surrounding context. By withholding $I$, the backward path must reconstruct the addressed problem from the patch rather than repeating the task originally given to the agent.

\paragraph{Patch interpretation.}
RETRACE first determines the functional meaning of the candidate patch. It examines the modified program elements, their relationships to the surrounding implementation, and the conditions under which the affected behavior is reached. It then identifies the behavior replaced or modified by the patch, the behavior introduced by the change, and the inputs, states, or execution paths to which the change applies. The objective is not to summarize the diff line by line, but to recover the behavior expressed by the implementation.

A candidate patch may contain several edits that serve different roles. Some directly alter the target behavior, while others update callers, propagate state, handle boundary cases, or preserve compatibility. RETRACE interprets these edits together to determine the overall change implemented by the patch rather than treating each modified location independently.

\paragraph{Trajectory-grounded interpretation.}
The patch alone may leave aspects of the implemented change ambiguous. For example, the purpose of a new condition may depend on an execution path inspected earlier, while the role of a modified argument may become clear only from the callers examined before the edit. RETRACE therefore uses $\tau_{\mathrm{b}}$ to recover the repository context surrounding the implementation, including the relevant program relationships, observations, and decisions made while constructing the patch. The patch serves as the primary evidence of the implemented behavior, while the trajectory is used to disambiguate its context. When an intention stated in trajectory conflicts with the completed change, RETRACE grounds the reconstruction in what the patch actually implements. This prevents an earlier repair intention from masking missing behavior, unintended scope, or modifications addressing a different problem.

\paragraph{Problem reconstruction.}
RETRACE combines the patch interpretation and trajectory context into the reconstructed problem $\widehat{I}$. The reconstruction describes the behavior affected by the patch, the conditions under which it arises, the behavior before and after the change, the affected components, and the apparent scope and intent of the implementation. Details unsupported by the patch or its surrounding context are not inferred from the original issue.

The reconstructed $\widehat{I}$ is not a judgment that the candidate patch is correct. It is an implementation-grounded account of the problem the patch appears to address, which RETRACE subsequently compares with the original issue $I$.

\subsection{Forward--Backward Reconciliation}
\label{sec}

Given the forward rationale $\phi$ and the backward reconstruction $\widehat{I}$, RETRACE determines whether the inconsistency originates from the reasoning, the implementation, or both. It first compares $\widehat{I}$ with the original issue $I$:
\[(v, d) = A(I, \widehat{I}),\]
where $v$ is the alignment verdict and $d$ records the agreements, omissions, and conflicts between the two problem descriptions. Reconciliation then jointly considers $I$, $\phi$, $p$, $\widehat{I}$, and $(v, d)$ to decide whether the patch should be submitted or revised. When a discrepancy is detected, it attributes the source and produces a targeted revision plan. The agent revises the patch accordingly, and RETRACE repeats backward reconstruction and reconciliation until the patch is aligned or the revision budget is exhausted.

\paragraph{Issue alignment.}
RETRACE first compares the reconstructed problem $\widehat{I}$ with the original issue:
\[
(v,d)=A(I,\widehat{I}),
\]
where
\[
v\in
{
\textsc{Same},
\textsc{Partial},
\textsc{Different}
}
\]
is the alignment verdict and $d$ is a structured disagreement record.

The comparison considers the reported behavior, triggering conditions, behavior before and after the expected change, affected scope, and repair intent. It evaluates their semantic agreement rather than textual similarity. Two descriptions may use different terminology while referring to the same behavior and scope. Conversely, descriptions may mention the same functions or concepts while differing in the conditions or behavior they address.

A \textsc{Same} verdict indicates that $\widehat{I}$ and $I$ describe the same problem with compatible behavior and scope. A \textsc{Partial} verdict indicates that the patch addresses some aspects of the issue but omits, narrows, or extends others. A \textsc{Different} verdict indicates that the problem expressed by the patch is materially different from the original issue. The disagreement record $d$ identifies the specific agreements, omissions, and conflicts underlying the verdict.

\paragraph{Patch--reasoning consistency.}
The alignment verdict identifies whether the patch appears to address the original issue, but it does not explain how any mismatch arose. RETRACE therefore also examines the candidate patch $p$ against the forward reconstruction $\phi=(E,r,s)$. It checks whether the patch implements the intended repair $s$, preserves the behavior and scope established by $r$, and avoids changes unsupported by the recovered evidence $E$.

This comparison considers the patch as a whole. Modifying the expected program location is insufficient when the implementation omits required behavior, handles only a subset of the relevant cases, or introduces changes outside the intended scope. Similarly, a patch may align with part of the issue while revealing that the intended repair itself was based on missing or incorrectly connected evidence.

\paragraph{Revision diagnosis.}
Using the issue-alignment result and the patch--reasoning comparison, RETRACE identifies where the observed inconsistency appears. An implementation discrepancy occurs when the reconstructed reasoning supports the intended repair, but the patch realizes it incompletely or differently. A reasoning discrepancy occurs when the evidence and reconstructed chain do not adequately support the intended repair, including cases in which relationships exposed during implementation were omitted from the earlier reasoning. The two discrepancies may also occur together.

This diagnosis does not claim to recover the true cause of failure without ground truth. Instead, it determines which parts of the available reasoning and implementation should be revisited in the next repair attempt.

\paragraph{Reconciliation decision.}
Given $I$, $\phi$, $p$, $\widehat{I}$, and $(v,d)$, RETRACE derives
\[
(a,s')=R\bigl(I,\phi,p,\widehat{I},v,d\bigr),
\]
\noindent\small{$a\in
{
\textsc{Submit},
\textsc{Revise-Patch},
\textsc{Revisit-Reasoning},
\textsc{Revise-Both}
}$}
is the reconciliation decision and
$s'=\langle s'_1,\ldots,s'_m\rangle$
is an ordered revision plan when further repair is required.

RETRACE selects \textsc{Submit} when the patch implements the intended repair and $\widehat{I}$ matches the original issue. It selects \textsc{Revise-Patch} when the reasoning provides an adequate basis for the repair but the implementation does not fully realize it. It selects \textsc{Revisit-Reasoning} when the reconstructed chain contains missing or unsupported connections that should be reconsidered before further editing. It selects \textsc{Revise-Both} when the reasoning and implementation both require revision.

Each step $s'_i$ identifies the code location or repair component to reconsider, the discrepancy motivating the revision, and the intended behavioral effect. The plan may direct the agent to recover missing evidence, reconnect an unsupported reasoning step, implement omitted behavior, narrow an overly broad change, or redirect an edit toward the relevant program component.

\paragraph{Revision through verification feedback.}
RETRACE renders the reconciliation result as a verification note containing the alignment verdict, the disagreement record, the identified reasoning or implementation discrepancies, the selected decision, and the ordered revision plan. This note is appended to the agent's working context, after which the underlying agent resumes its existing repository interaction and editing process. RETRACE therefore guides revision without replacing the agent's tools or edit mechanism.

After the agent produces an updated patch, RETRACE repeats backward reconstruction and reconciliation using the revised implementation. The forward reconstruction is revisited only when the reconciliation decision identifies a reasoning discrepancy. The process terminates when RETRACE selects \textsc{Submit} or when the configured revision budget is exhausted.

\section{Experiments}
\label{sec:exp}
% ============================================================================

\subsection{Experimental Settings}

\paragraph{Base models.}
We evaluate RETRACE with two lower-cost yet strong models: \textbf{GPT-5 mini}~\cite{openai2025gpt5} and \textbf{MiniMax M2.5}~\cite{minimax2025m25}. Using models from different providers allows us to examine whether the results hold across multiple models while keeping evaluation costs manageable. We use greedy decoding to reduce run-to-run variation and report one run per configuration. 
%[TODO: confirm temperature/greedy and single-run vs.\ mean.]

\paragraph{Scaffold and implementation.}
% \peter{there is open hand right? mention it here. say we do miniswe agent, and do openhand to if this approach works on another agent scaffold}We build RETRACE on top of mini-SWE-agent \cite{yang2024sweagent}, a minimalist agent scaffold whose core logic is approximately 100 lines of Python.  \peter{too much detail}The agent interacts with the repository exclusively through bash commands, without relying on custom tools or the model's tool-calling interface. RETRACE adds only the forward reasoning, backward inference, and reconciliation stages around the existing edit loop, making it a drop-in addition rather than a new scaffold.  All added components use the same backbone model as the underlying agent.
We build RETRACE on the scaffold \textbf{ mini-SWE-agent}~\cite{yang2024sweagent}, a minimalist agent that interacts with the repository through bash commands. To test generalizability, we extend RETRACE on another scaffold, \textbf{OpenHands}~\cite{openhands}, which adopts a CodeAct paradigm with a different tool interface and agent architecture. RETRACE adds the forward reasoning, backward reconstruction, and reconciliation stages around each scaffold's existing edit loop without modifying its internal logic. All added components use the same backbone model as the underlying agent.

\paragraph{Benchmark and metrics.}
We conduct our experiments on \textbf{SWE-bench Verified}~\cite{jimenez2024swebench}, a human-validated subset of 500 real-world GitHub issues from SWE-bench, each equipped with fail-to-pass and pass-to-pass tests. Following the previous studies~\cite{xia2024agentless}, we evaluate the efficacy of RETRACE using  \textbf{Pass@1}, the percentage of issues for which the single submitted patch passes the complete test suite.

\paragraph{Baselines.}
Our baseline is the underlying mini-SWE-agent with the same model. We also compare against \textbf{Self-Refine}~\cite{madaan2023selfrefine}, which reviews the candidate patch against the original issue and returns revision feedback, using up to three revision iterations with a comparable inference budget.

% ----------------------------------------------------------------------------
\subsection{Experiment and Results}

\begin{table}[t]
\centering
\resizebox{\columnwidth}{!}{%
\begin{tabular}{llccc}
\toprule
\textbf{Model} & \textbf{Method} & \textbf{Resolved(\#)} & \textbf{Pass@1(\%)} & \textbf{$\Delta$} \\
\midrule
 & mini-SWE-agent              & 281/500 & 56.2 & --- \\
 
GPT-5 mini &  Self-Refine & 274/500 & 54.8 & -1.4 \\
 &   RETRACE              & \textbf{316/500} & \textbf{63.2} & \textbf{+7.0} \\
\midrule
 & mini-SWE-agent              & 379/500 & 75.8 & --- \\
 MiniMax M2.5 &  Self-Refine & 370/500 & 74.0 & -1.8 \\
 & RETRACE              & \textbf{397/500} & \textbf{79.4} & \textbf{+3.6} \\
\bottomrule
\end{tabular}}
\caption{Main results on SWE-bench Verified (500 issues).}
\label{tab:main}
\end{table}

\paragraph{Performance of RETRACE.} 
Table~\ref{tab:main} compares RETRACE with the underlying mini-SWE-agent and Self-Refine on the full SWE-bench Verified benchmark. RETRACE improves Pass@1 with both backbone models, from 56.2\% to 63.2\% with GPT-5 mini and from 75.8\% to 79.4\% with MiniMax M2.5, corresponding to gains of 7.0 and 3.6 percentage points. These results show that RETRACE improves issue resolution across two backbones with different architectures and baseline performance. 

Table~\ref{tab:main} also shows that Self-Refine~\cite{madaan2023selfrefine}, despite using a comparable inference %budget, does not improve either baseline. It reduces \% Resolved by 1.4 percentage points with GPT-5 mini and 1.8 points with MiniMax M2.5. 
despite a comparable inference budget, does not improve either baseline, reducing Pass@1 by 1.4 and 1.8 percentage points for GPT-5 mini and MiniMax M2.5, respectively. This suggests that iterative review alone is insufficient. Unlike Self-Refine, RETRACE reconstructs both repair reasoning and the problem implied by the completed patch, highlighting the benefit of bidirectional validation.
\begin{table}[t]
\small
\centering
\begin{tabular}{lcc}
\toprule
\textbf{Configuration} & \textbf{Resolved(\#)} & \textbf{Pass@1(\%)} \\
\midrule
Baseline                       & 60/120          & 50.0 \\
RETRACE $_{Forward-only}$
    & 68/120          & 56.7 \\
RETRACE $_{Backward-only}$   & 68/120          & 56.7 \\
\textbf{RETRACE (full)}                 & \textbf{73/120} & \textbf{60.8} \\
\bottomrule
\end{tabular}
\caption{Component ablation on a random 120-issue subset of
SWE-bench Verified (GPT-5-mini).}
\label{tab:ablation}
\end{table}

\paragraph{Ablation Studies.} 
% To isolate the contribution of each stage, we evaluate four configurations on a random 120 issue subset of SWE-bench Verified
% using GPT-5-mini.  (1)~\textit{\textbf{Baseline}}: mini-SWE-agent without any RETRACE intervention.  
% (2)~\textit{\textbf{RETRACE w/o backward spec. infer.}}: 
% Only forward reasoning is applied between search and editing (pruning irrelevant context and injecting the repair plan), but the agent submits the resulting patch directly without backward verification. 
% This isolates the contribution of giving the agent a structured direction before it begins editing. 
% (3)~\textit{\textbf{RETRACE w/o forward reasoning}}: Only backward
% specification inference is  introduced after the agent submits its patch, which reconstructs the problem specification from the generated patch and decides whether to revise directly based on comparison with the original issue.
% This isolates the contribution of post-editing verification when the agent has received no prior guidance.
% (4)~\textit{\textbf{RETRACE (full)}}: all three stages are applied.

To isolate the contribution of each stage, we evaluate four configurations on a random 120-issue subset of SWE-bench Verified using GPT-5-mini (Table~\ref{tab:ablation}):
(1)~\emph{\textbf{Baseline}}: mini-SWE-agent without any RETRACE intervention. (2)~\emph{\textbf{Forward only}}: forward reasoning is applied. (3)~\emph{\textbf{Backward only}}: only backward specification inference is applied. (4)~\emph{\textbf{RETRACE (full)}}: all stages are applied.  We observe that \textbf{\textit{each stage contributes to RETRACE's performance, and their
combination leads to further improvement.}} Removing either stage reduces Pass@1
from 60.8\% to 56.7\%, while removing both returns to the baseline (50.0\%).
Although the two single-stage variants achieve the same aggregate Pass@1, they
rescue different sets of baseline failures, indicating that forward reasoning and
backward specification inference address distinct error types. When all stages
are combined, RETRACE achieves the highest Pass@1 (60.8\%), confirming that the complementary effects of both stages.
%stages complement and reinforce one another.
% : forward reasoning corrects misguided editing decisions by pruning irrelevant context and injecting a repair plan, backward specification inference catches patches that diverge from the reported problem, and reconciliation synthesizes both signals into a targeted revision.

\begin{figure}[t]
\centering
\includegraphics[width=0.5\columnwidth]{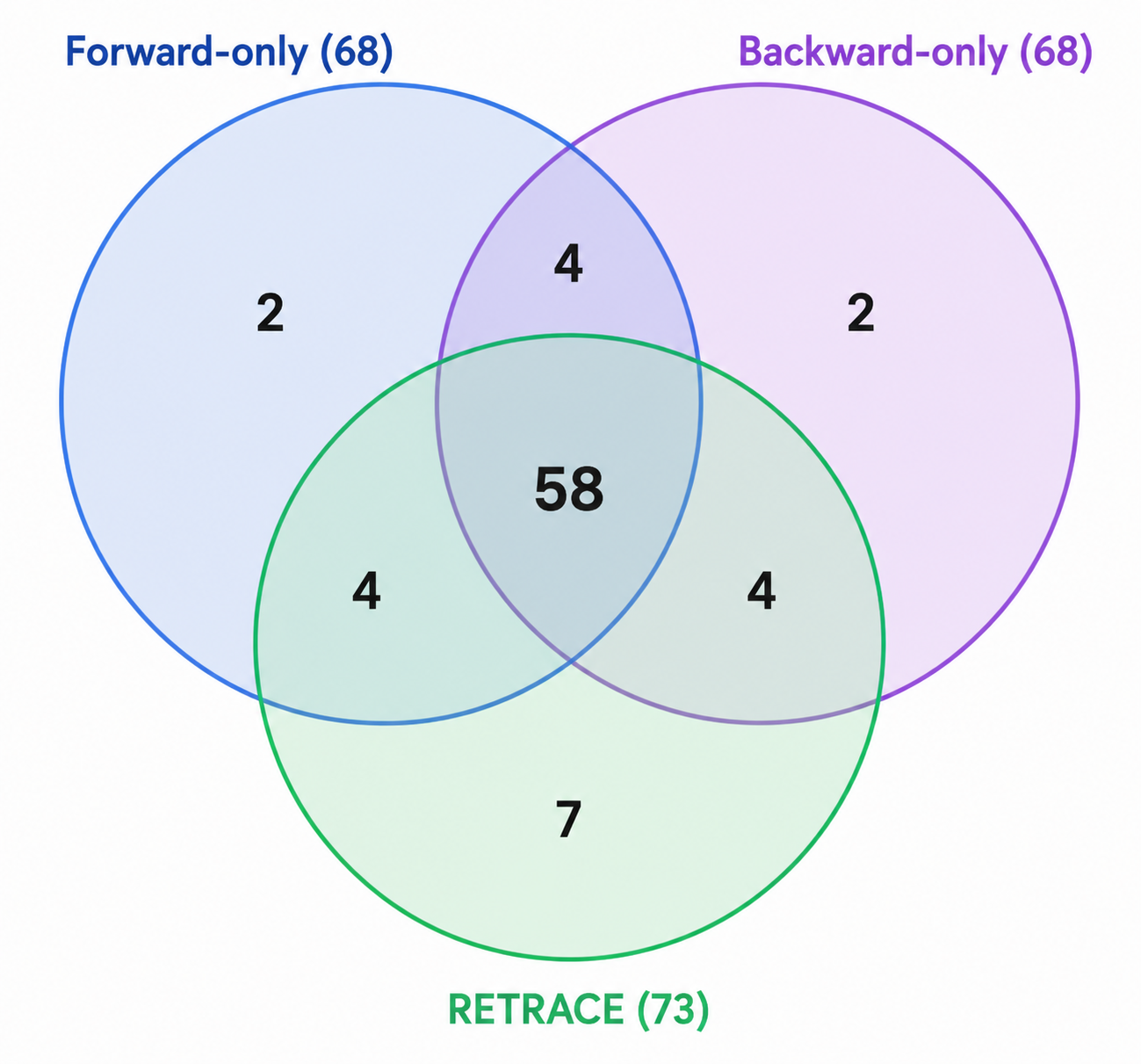}
\caption{Overlap of resolved issues among three configurations on the
120-issue subset (GPT-5-mini).}
\label{fig:overlap}
\end{figure}

We further analyze the overlap of resolved issues across configurations. As shown in Figure~\ref{fig:overlap} Forward-only and Backward-only each resolve 68 of the 120 issues, but each resolves 6 issues that the other misses: forward reasoning corrects flawed editing decisions by reorganizing context before the patch is produced, while backward specification inference catches patches whose effect diverges from the reported problem after the fact. Full RETRACE retains 66 issues resolved by each single-stage variant while resolving 7 additional issues that neither variant resolves alone. Reconciliation thus not only consolidates the gains of both paths but also exposes errors that neither direction detects in isolation.

% strongest result. Removing either one degrades performance: RETRACE w/o backward verif.\ reduces \% Resolved from 60.8\% to 56.7\%; RETRACE w/o forward planning also reduces it to 56.7\%.  Although the two ablations yield the same resolve rate, the sets of resolved instances differ, indicating that the two stages address distinct failure modes.  Removing both returns to the baseline of 50.0\%. Forward reasoning targets failures that arise before editing: the agent may attend to irrelevant context or lack a clear direction, leading to a misguided patch. 
% \shinhwei{the approach section used "backward specification inference" but here use "backward verification". use consistent term}
% Backward specification inference targets failures that arise after editing: the patch may drift from the original problem, a divergence that is invisible without an independent check.  The full integration yields the strongest result because reconciliation can attribute disagreement to reasoning, implementation, or both, producing a more targeted revision signal than either stage provides alone.
% ----------------------------------------------------------------------------
% \subsection{Scaffold Generalizability}

\begin{table}[t]
\small
\centering
\begin{tabular}{llccc}
\toprule
\textbf{Model} & \textbf{Method} & \textbf{Resolved} & \textbf{Pass@1} & \textbf{$\Delta$} \\
\midrule

\multirow{3}{*}{ GPT-5-mini}
 & OpenHands   & 45/120 & 37.5 & --- \\
 & Self-Refine & 64/120 & 53.3 & +15.8\\
 & RETRACE  & \textbf{68/120} & \textbf{56.7} & \textbf{+19.2} \\
\midrule
\multirow{3}{*}{ MiniMax M2.5}
 & OpenHands   & 75/120 & 62.5 & --- \\
  & Self-Refine & 77/120& 64.2& +1.7\\
 & RETRACE  & \textbf{84/120} & \textbf{70.0} & \textbf{+7.5} \\
\bottomrule
\end{tabular}
\caption{Scaffold generalizability on the same 120-issue subset.
RETRACE is applied as a drop-in addition to each scaffold without
modifying its tool interface.  mini-SWE-agent results use GPT-5-mini.}
\label{tab:generalize}
\end{table}

\paragraph{Generalizability to other scaffolds.}
We further evaluate RETRACE's ability to generalize to other code agent scaffold OpenHands \cite{openhands}. Table~\ref{tab:generalize} reports the results of extending RETRACE to OpenHands on the same 120-issue subset. \textbf{\textit{RETRACE consistently outperforms both the vanilla OpenHands and Self-Refine across both base models, confirming that its improvement is not tied to a specific agent framework.}} With GPT-5-mini, RETRACE raises Pass@1 from 37.5\% to 56.7\%. With MiniMax-2.5, the Pass@1 also increase from 62.5\% to 70.0\%. Self-Refine also improves over the OpenHands baseline on both models, unlike on mini-SWE-agent, where it slightly degrades performance, suggesting that OpenHands benefits from additional post-edit reasoning in general. However, RETRACE outperforms Self-Refine in both settings models, indicating that the gain comes not only from additional inference but from the independent verification signal that
reconciliation provides.

% ----------------------------------------------------------------------------
% \subsection{Cost Analysis}
% \label{sec:cost}

% \begin{table}[t]
% \centering
% \small
% \begin{tabular}{llccc}
% \toprule
% \textbf{Model} & \textbf{Method} & \textbf{Input} & \textbf{Output} & \textbf{Ratio} \\
% \midrule
% \multirow{2}{*}{GPT-5 mini}
%  & Baseline & 230K & 3.9K & 1.00$\times$ \\
%  & RETRACE  & 405K & 7.6K & 1.68$\times$ \\
% \midrule
% \multirow{2}{*}{MiniMax M2.5}
%  & Baseline & \textit{TBD} & \textit{TBD} & 1.00$\times$ \\
%  & RETRACE  & \textit{TBD} & \textit{TBD} & \textit{TBD}$\times$ \\
% \bottomrule
% \end{tabular}
% \caption{Average per-issue token usage on SWE-bench Verified
% (500 issues).  Ratio is relative to each model's own baseline.}
% \label{tab:cost}
% \end{table}

\begin{table}[t]
\small
\centering
\begin{tabular}{llccc}
\toprule
\textbf{Model} & \textbf{Method} & \textbf{Input}  & \textbf{CacheHit} & \textbf{Cost} \\
\midrule
\multirow{2}{*}{GPT-5 mini}
 & mini-SWE-agent & 26.3K  & 45.6\% & \$0.07 \\
 & RETRACE  & 60.6K  & 90.8\%  & \$0.06 \\
\midrule
\multirow{2}{*}{MiniMax M2.5}
 & Baseline & 123.3K  & 94.6\%  & 0.07 \\
 & RETRACE  & 78.0K  & 96.7\%  & 0.05 \\
\bottomrule
\end{tabular}
\caption{Average per-issue token usage and cost on SWE-bench
Verified.}
\label{tab:cost}
\end{table}

% \begin{table}[t]
% \centering
% \small
% \begin{tabular}{llcccc}
% \toprule
% \textbf{Model} & \textbf{Method} & \textbf{Avg.\ input tok.} & \textbf{Avg.\ output tok.} & \textbf{Total tok.} & \textbf{Ratio} \\
% \midrule
% \multirow{2}{*}{GPT-5-mini}
%  & Baseline & 230K & 3.9K & 234K & 1.00$\times$ \\
%  & RETRACE  & 405K & 7.6K & 413K & 1.68$\times$ \\
% \midrule
% \multirow{2}{*}{MiniMax-2.5}
%  & Baseline & \textit{TBD} & \textit{TBD} & \textit{TBD} & 1.00$\times$ \\
%  & RETRACE  & \textit{TBD} & \textit{TBD} & \textit{TBD} & \textit{TBD}$\times$ \\
% \bottomrule
% \end{tabular}
% \caption{Per-issue token usage on SWE-bench Verified (500 issues).
% RETRACE adds forward reasoning, backward inference, specification
% alignment, and reconciliation calls on top of the baseline.}
% \label{tab:cost}
% \end{table}

\paragraph{Cost analysis.}
Table~\ref{tab:cost} reports average per-issue token usage and cost. Although RETRACE introduces additional reasoning stages, its per-issue cost remains comparable to or below the baseline on both backbones. Two factors contribute. First, the forward, backward, and reconciliation stages share the issue description and agent trajectory as input, yielding high prompt-cache hit rates (90.88\% with GPT-5-mini, 96.73\% with MiniMax-2.5). Second, targeted verification feedback reduces the agent's exploratory output, as reflected in the lower output-token count with GPT-5-mini (1.2K vs.\ 4.2K). The overall cost ratio remains below $1.0\times$ on both backbones, indicating that bidirectional validation does not impose a practical cost overhead.

%[TODO: fill in MiniMax-2.5 token numbers.]

% % ----------------------------------------------------------------------------
% \subsection{Mechanism Analysis}
% \label{sec:mechanism}

% %[TODO: fill in after extracting statistics from logs.]

% \paragraph{Alignment verdict calibration.}
% %How accurate is the specification alignment verdict?  We manually
% %inspect [TODO: N] issues and compare the model's
% %same/partial/different verdict against human judgment.
% %[TODO: report accuracy, confusion matrix, or agreement rate.]

% \paragraph{Rescue and regression analysis.}
% %Among the issues that RETRACE resolves but the baseline does not
% %(rescues), and vice versa (regressions):
% %[TODO: report rescue count, regression count, net gain.]
% %[TODO: break down rescues by which stage contributed: forward-only
% %rescue, backward-only rescue, both required.]

% % ----------------------------------------------------------------------------
% \subsection{Qualitative Example}
% \label{sec:qualitative}

% %[TODO: add a case study showing the full RETRACE loop on one issue.
% %Include: (1) the original issue, (2) the baseline's incorrect patch,
% %(3) the forward repair rationale, (4) the backward-inferred
% %specification, (5) the alignment verdict and disagreement record,
% %(6) the reconciliation direction and revision plan, (7) the
% %corrected patch.]
\section{Limitations}

Our evaluation is conducted on SWE-bench Verified, a human-validated subset of 500 real-world GitHub issues that primarily targets Python projects, which may restrict generalization to other languages or domains. In addition, RETRACE's backward verification compares a reconstructed problem against the original issue report, so its signal depends on issue quality. Vague or underspecified reports weaken the alignment verdict, though the human-validated nature of SWE-bench Verified bounds this effect in practice. Both the forward reconstruction and backward specification inference rely on the same backbone model that generated the patch, and their effectiveness may vary with the underlying model's reasoning reliability, using a heterogeneous verifier is a natural extension that our model-agnostic design already accommodates.
\section{Conclusion}

We introduced RETRACE, a framework that validates software patches through bidirectional reasoning. After an agent produces a candidate patch, RETRACE reconstructs the repair rationale from the issue and trajectory in the forward direction, and reconstructs the addressed problem from the patch and trajectory in the backward direction, withholding the original issue to ensure independence. A reconciliation step then compares the two reconstructions to determine whether the patch should be submitted or revised. On SWE-bench Verified, RETRACE improves Pass@1 from 56.2\% to 63.2\% with GPT-5-mini and from 75.8\% to 79.4\% with MiniMax-2.5, with consistent gains on a second scaffold. Ablations confirm that the forward and backward paths address complementary failure modes, and that reconciliation produces repair signals beyond what either path provides alone. These results suggest that independent reconstruction-based verification offers a practical alternative to test-based or self-review signals for improving the reliability of autonomous issue resolution.

\end{document}